\input{psfig}
\documentstyle[onecolumn,doublespacing]{mn}

\newcommand{\be}{\begin{equation}}
\newcommand{\ee}{\end{equation}}
\newcommand{\bea}{\begin{eqnarray}}
\newcommand{\eea}{\end{eqnarray}}
\newcommand{\bc}{\begin{center}}

\newcommand{\ec}{\end{center}}
\def\spose#1{\hbox to 0pt{#1\hss}}
\newcommand{\lta}{\mathrel{\spose{\lower 3pt\hbox{$\mathchar"218$}}
     \raise 2.0pt\hbox{$\mathchar"13C$}}}
\newcommand{\gta}{\mathrel{\spose{\lower 3pt\hbox{$\mathchar"218$}}
     \raise 2.0pt\hbox{$\mathchar"13E$}}}

\def\H0{$H_0= 100~h~$km\,s$^{-1}$\,Mpc$^{-1}$}

\newif\ifAMStwofonts
\ifoldfss
  \ifCUPmtlplainloaded \else
    \NewTextAlphabet{textbfit} {cmbxti10} {}
    \NewTextAlphabet{textbfss} {cmssbx10} {}
    \NewMathAlphabet{mathbfit} {cmbxti10} {} % for math mode
    \NewMathAlphabet{mathbfss} {cmssbx10} {} %  "   "    "
  \fi
  \ifAMStwofonts
    \ifCUPmtlplainloaded \else
      \NewSymbolFont{upmath} {eurm10}
      \NewSymbolFont{AMSa} {msam10}
      \NewMathSymbol{\upi}     {0}{upmath}{19}
      \NewMathSymbol{\umu}     {0}{upmath}{16}
      \NewMathSymbol{\upartial}{0}{upmath}{40}
      \NewMathSymbol{\leqslant}{3}{AMSa}{36}
      \NewMathSymbol{\geqslant}{3}{AMSa}{3E}

      \let\leq=\leqslant \let\le=\leqslant
      \let\geq=\geqslant \let\ge=\geqslant
    \fi
  \fi
\fi % End of OFSS

\ifnfssone
  \newmathalphabet{\mathit}
  \addtoversion{normal}{\mathit}{cmr}{m}{it}
  \addtoversion{bold}{\mathit}{cmr}{bx}{it}
  \newmathalphabet{\mathbfit} % math mode version of \textbfit{..}
  \addtoversion{normal}{\mathbfit}{cmr}{bx}{it}
  \addtoversion{bold}{\mathbfit}{cmr}{bx}{it}
  \newmathalphabet{\mathbfss} % math mode version of \textbfss{..}
  \addtoversion{normal}{\mathbfss}{cmss}{bx}{n}
  \addtoversion{bold}{\mathbfss}{cmss}{bx}{n}
  \ifAMStwofonts
    \ifCUPmtlplainloaded \else
and
your
      \UseAMStwoboldmath
      \makeatletter
      \new@mathgroup\upmath@group
      \define@mathgroup\mv@normal\upmath@group{eur}{m}{n}
      \define@mathgroup\mv@bold\upmath@group{eur}{b}{n}
      \edef\UPM{\hexnumber\upmath@group}
      \new@mathgroup\amsa@group
      \define@mathgroup\mv@normal\amsa@group{msa}{m}{n}
      \define@mathgroup\mv@bold\amsa@group{msa}{m}{n}
      \edef\AMSa{\hexnumber\amsa@group}
      \makeatother
      \mathchardef\upi="0\UPM19
      \mathchardef\umu="0\UPM16
      \mathchardef\upartial="0\UPM40
      \mathchardef\leqslant="3\AMSa36
      \mathchardef\geqslant="3\AMSa3E

      \let\leq=\leqslant \let\le=\leqslant
      \let\geq=\geqslant \let\ge=\geqslant
    \fi
  \fi
\fi % End of NFSS release 1

\ifnfsstwo
  \DeclareMathAlphabet{\mathbfit}{OT1}{cmr}{bx}{it}
  \SetMathAlphabet\mathbfit{bold}{OT1}{cmr}{bx}{it}
  \DeclareMathAlphabet{\mathbfss}{OT1}{cmss}{bx}{n}
  \SetMathAlphabet\mathbfss{bold}{OT1}{cmss}{bx}{n}

  \ifAMStwofonts
    \ifCUPmtlplainloaded \else
      \DeclareSymbolFont{UPM}{U}{eur}{m}{n}
      \SetSymbolFont{UPM}{bold}{U}{eur}{b}{n}
      \DeclareSymbolFont{AMSa}{U}{msa}{m}{n}
      \DeclareMathSymbol{\upi}{0}{UPM}{"19}
      \DeclareMathSymbol{\umu}{0}{UPM}{"16}
      \DeclareMathSymbol{\upartial}{0}{UPM}{"40}
      \DeclareMathSymbol{\leqslant}{3}{AMSa}{"36}
      \DeclareMathSymbol{\geqslant}{3}{AMSa}{"3E}

      \let\leq=\leqslant \let\le=\leqslant
      \let\geq=\geqslant \let\ge=\geqslant
    \fi
  \fi
\fi % End of NFSS release 2

\ifCUPmtlplainloaded \else
  \ifAMStwofonts \else % If no AMS fonts
    \def\upi{\pi}
    \def\umu{\mu}
    \def\upartial{\partial}
  \fi
\fi

\title{Migration of giant planets in planetesimal discs.}
\author[A.Del Popolo et al.]
  {A. Del Popolo,$^1$$^,$$^2$$^,$$^3$ M. Gambera,$^1$ E. Nihal Ercan$^3$\\
  $^1$ Dipartimento di Matematica, Universit\`{a} Statale di Bergamo,
  Piazza Rosate, 2 - I 24129 Bergamo, ITALY \\
     $^2$ Feza G\"ursey Institute, P.O. Box 6 \c Cengelk\"oy, Istanbul,
     Turkey\\
     $3$  Bo$\breve{g}azi$\c{c}i University, Physics Department,
     80815 Bebek, Istanbul, Turkey
}
\date{Accepted ???
      Received 2000 July 24;
      in original form ???}

\pagerange{\pageref{firstpage}--\pageref{lastpage}}
\pubyear{2000}

\begin{document}

\maketitle

\label{firstpage}

\begin{abstract}
Planets orbiting a planetesimal circumstellar disc can migrate
inward from their initial positions because of dynamical friction
between
planets and planetesimals. The migration rate depends on the
disc mass and on its time evolution. Planets that are embedded in
long-lived
planetesimal discs, having total mass of $10^{-4}-0.01 M_{\odot}$,
can migrate inward a large distance
and can survive only if the inner disc is truncated or because
of tidal interaction with the star. In this case the semi-major axis,
$a$,
of the planetary orbit is less than $0.1 {\rm AU}$. Orbits with larger
$a$ are obtained
for smaller value of the disc mass or for a rapid evolution (depletion)
of the disc. This model
may explain several of the orbital features of the giant planets
that were discovered in last years orbiting nearby stars
as well as the metallicity
enhancement found in several stars associated with short-period planets.
\end{abstract}


%\newpage

\section{Introduction}

\indent According to the most popular theory on the formation of giant
planets in the solar system, planets were formed by accumulation of
solid
cores (Safronov 1969; Wetherill \& Stewart 1989; Aarseth et al. 1993),
known as planetesimals, in a gaseous disc centered around the
sun. When the core mass increases above $10 M_{\odot}$,
it begins a rapid accretion phase (Mizuno 1980; Bodenheimer \& Pollack
1986)
in which the protoplanet can capture a gas envelope from the
protoplanetary
disc leading to the formation of a giant planet (Pollack et al. 1996).
Jupiter-mass planets may require most of the lifetime of the disc to
accrete ($10^6$-$10^7$ yr) (Zuckerman et al. 1995; 
Pollack et al. 1996). \\
\indent Protostellar discs around young stellar objects that have properties
similar to that supposed for the solar nebula are common: between 25 to 75\%
of young stellar objects in the Orion nebula seem to have discs (Prosser
et al. 1994; McCaughrean \& Stauffer 1994) with mass 
$10^{-3} M_{\odot}<M_{\rm d} <10^{-1} M_{\odot}$ and size $40 \pm 20$
AU (Beckwith \& Sargent 1996). Moreover recently several planetary 
companion orbiting extra-solar stars were discovered. The extrasolar planets
census, updated at October 2000, gives 58 planets. For reason of space, 
we report only a small number of them: the companions orbiting
%%PSR B1257+12 (Wolszczan \& Frail
%%1992),
51 Peg (Mayor \& Queloz 1995), $\tau$ Boo(Marcy et al. 1997 - San 
Francisco University Team hereafter SFSU), $v$ And (SFSU), $\rho^1$ Cnc 
(SFSU), $\rho$ CrB (Noyes et al. 1997 - AFOE team), HD 114762,
70 Vir, 16 Cyg, 47 UMa (Butler \& Marcy 1996). In the above list, with the
exception of 47 Uma, the new planets are all at distances $<1 $AU.
In recent years has been discovered other extrasolar planets orbiting at distances $>1 {\rm AU}$ from the central star:\\
$\epsilon$ Eridani, HD210277, HD 82943, 14 Her, HD 190228, HD 222582,
HD10697, HD 29587, representing only 15\% of all the planets.
Three planets (51 Peg, $\tau$ Boo, $v$ And) are in extremely tight circular 
orbits with periods of a few days, two planets ($\rho^1$ Cnc and $\rho$ 
CrB)have circular orbits with periods of order tens of days and  three 
planets with wider orbits (16 Cyg B, 70 Vir and HD 114762) have very 
large eccentricities. The properties of these planets, most of which 
are Jupiter-mass objects, are difficult to explain using the quoted 
standard model for planet formation (Lissauer 1993; Boss 1995).
This standard model predicts nearly circular planetary orbits, and giant
planets distances $\geq 1$ AU from the central star, distance at which 
the temperature in the protostellar nebula is low enough for icy materials 
to condense (Boss 1995, 1996; Wuchterl 1993, 1996). Standard disc models 
show that at 0.05 AU, 
%%(region in which were found some of the planets), 
the temperature is about 2,000 K, which is too hot for the existence of 
any small solid particles. Moreover the ice condensation radius does not 
depend strongly on stellar mass, so that it does not move inward rapidly 
as the stellar mass decreases. For star masses, $M_{\ast}=1$, 0.5, 0.1 
$M_{\odot}$, the ice condensation radius moves inward from $\simeq 6$ to 
$\simeq 4.5$ AU(Boss 1995). Another problem with the {\it in situ} formation 
of a planetary companion is that even though the present evaporation rate is 
negligible, this effect would have been of major importance in the past. In 
fact during the early history of a planet, its radius was a factor  ten 
larger than the present value, implying that the escape speed was much less 
than its present value. Hence evaporation mechanisms and ablation by the 
stellar wind might prevent its formation. The question that arises is: if 
such massive planets cannot form at the actual locations, how did they reach
their actual position ? Four mechanisms have been proposed to explain the 
quoted dilemma.\\
\indent The first mechanism consists of a secular interaction with a distant 
binary companion (Holman et al. 1997; Mazeh et al. 1996; Kiseleva \& 
Eggleton 1997; Eggleton \& Kiseleva 1997). While this mechanism
can also produce significant eccentricities for the longer period extrasolar 
planets it is unable to explain objects like 51 Peg. In fact, 51 Peg has 
been extensively searched for a binary companion (Marcy et al. 1997), but 
none has been found. Consequently, in the particular case of 51 Peg,
this mechanism is not responsible for the orbital decay.\\
\indent The second possible mechanism proposed to explain short period
planets is dissipation in the protostellar nebula.
%%(Goldreich \& Tremaine 1979, 1980; Ward 1986; Lin et al. 1996; Ward 
%%1997).
Tidal interaction between a massive planet and a circumstellar disc gives 
rise to an angular momentum transfer between the disc and the planet
(Goldreich \& Tremaine 1979, 1980; Ward 1986; Lin et al. 1996; Ward 1997). 
The planet's motion in the disc excites density waves both interior and
exterior to the planet. A torque originates from the attraction of
the protoplanet for these non-axisymmetric density perturbations
(Goldreich \& Tremaine 1980). Density wave torques repel material on either 
side of the protoplanet's orbit and attempt to open a gap in the disc,
whose size depends on the viscosity of the disc and inversely on the
planet's mass (Lin \& Papaloizou 1986; Takeuchi et al. 1986)
(note that exists a minimum mass
for gap opening, which is of the order of magnitude of
Jovian planets mass, which prevents the nonsense of an infinite large
gap for a zero-mass planet). If gap formation is successful (for example in the case
of
a Jupiter mass planet),
the protoplanet becomes locked to the disc and must ultimately share its
fate (Ward 1982; Lin \& Papaloizou 1986, 1993). This mechanism is called
${\it type}$ II drift. The situation is different if the object is not
yet large enough to open and substain a gap. Also in this case
the protoplanet migrates inwards but with a time-scale even smaller than
that
of ${\it type}$ II drift (Ward 1997). This is called ${\it type}$ I
drift. In both cases, the rate of radial mobility of the planet,
with respect to the central star, is indicated with
the term 'drift velocity' (see Ward 1997)
(in some cases, see the case of Neptun below, the drift velocity can be
directed outwards).
Since, in this model, the time-scale of migration is $\simeq 10^5 \frac{M_{\rm
p}}{M_{\oplus}} {\rm yr}$
(Ward 1997),
the migration has to switch off at a critical moment, if the planet has
to stop close to the star without falling in it.
The movement of the planet might be halted by short-range tidal or magnetic
effects from the central star (Lin et al. 1996) (in any case,
as shown by Murray et al. (1998), it is
difficult to explain, by means of these stopping mechanisms,
planets with semi-major axes $a \ge 0.2$ AU). \\
%%
%%axes $a \ge 0.2$ AU . \\
%%
\indent A resonant interaction with a disc of planetesimals is another
possible source of orbital migration. In this model, planet migration starts 
when the surface density of planetesimals, $\Sigma$
%%is greater than some
satysfies the condition $\Sigma \ge \Sigma_{\rm c}$, being $\Sigma_{\rm c}$
a critical value for $\Sigma_{\rm c}$. The advantage of this mechanism is 
that the migration is halted naturally at short distances when the majority 
of perturbed planetesimals collide with the star. Moreover wide eccentric 
orbits can also be produced for planets more massive than 
$\simeq 3 M_{\rm J}$. However the model has some disavantages,
since the protoplanetary disc mass required for the migration of a
Jupiter-mass planet to $a \simeq 0.1$ AU is very large (Ford et al. 1999).\\
%%Moreover such a massive disc would probably produce more than one %%planet
%%which might have unknown effects on the model predictions.\\
\indent Finally, the fourth and last mechanism deals with dynamical
instabilities in a system of giant planets (Rasio \& Ford 1996).
The orbits of planets could become unstable if the orbital radii evolve 
secularly at different rates or if the masses increase significantly as the 
planets accrete their gaseous envelopes (Lissauer 1993). 
%%For example, according to Malhotra (1995), orbital migration 
%%has probably occurred in the 
%%outer solar system while Fernandez \& Ip (1984) showed that an %%%%%inwards 
%%migration of Jupiter of 0.1-0.2 AU can explain the depletion of the %%outer 
%%asteroid belt (Holman \& Murray 1996; Liou \& Malhotra 1997).
In this model, the gravitational interaction between two planets, during
evolution, (Gladman 1993; Chambers et al. 1996) can give rise to the ejection 
of one planet, leaving the other in eccentric orbit. 
The orbit of this can then circularize at an orbital separation of a few
stellar radii (Rasio et al. 1996) if the inner planet has
a sufficiently small pericenter distance. Simulations of multiple giant
planets systems, showed that successive mergers between two or more
planets can lead to the formation of a massive ($\ge 10 M_{\rm J}$) 
object in a wide eccentric orbit (Lin \& Ida 1997). 
While it is almost certain that this mechanism operates
in many systems with multiple planets, it is not clear
if they can reproduce the fraction of systems similar to 51 Peg 
observed. \\
In this paper we propose another model to explain the orbital parameters of extraterrestrial planets. The model is based on 
dynamical friction of a planet with a planetesimal disc. While 
the role of dynamical friction on the planetary accumulation
process has been studied in several papers (Stewart \& Kaula 1980; 
Horedt
1985; Stewart \& Wetherill 1988), very few papers have studied its role
on radial migration of planets or planetesimals in planetary discs
(see Melita \& Woolfson, 1996; Haghighipour 1999). This attitude is due 
to
the fact that, so far, many people have assumed a priori that radial
migration due to dynamical friction is much slower than the damping of velocity dispersion due to dynamical friction. Therefore most 
studies on dynamical friction were concerned only with damping of 
velocity dispersion (damping of the eccentricity, $e$, and 
inclination, $i$), adopting local coordinates. Analytical works by 
Stewart \& Wetherill (1988) and Ida (1990) adopted local 
coordinates. N-body simulation by Ida \& Makino (1992) adopted 
non-local coordinates, but did not investigate radial migration. 
Moreover most models of the planetesimal disc assume that, except 
for the influence of aerodynamic drag, which loses its effectiveness for planetesimals larger than a few kilometers, the primary cause 
of radial migration is mutual scattering (Hayashi et al. 1977 
and Wetherill 1990).
%%Radial migration was predicted by the controversial density
%%wave approach by Goldreich \& Tremaine (1979, 1980) but a caveat 
%%stated in that paper ("With our scant knowledge of the nebula we 
%%cannot be certain whether the interior or exterior torque is 
%%larger. Thus, we can estimate only the magnitude of the effect, 
%%not its sign") seems to have discouraged attempts to consider 
%%seriously radial migration of planetesimals 
%%in discs in further studies. \\
%%\indent This attitude has been reversed after the discovery of 
%%extra-solar
%%planets; most, as previously quoted, are Jupiter-mass objects in 
%%very tight circular orbits or in wider eccentric orbits. 
In order to calculate protoplanets migration, we apply the model 
introduced in Del Popolo et al. (1999) to study KBOs (Kuiper Belt 
Objects) migration, and we suppose that the gas in the 
disc is dissipated soon after 
the planet forms so that it has little effect on planet migration. 
We are particularly interested in studying the role of planetesimals
in planet migration and the dependence of migration on the disc 
mass and on its evolution.\\
%(Frizione per il gas?) \\
\indent The plan of the paper is the following: in Sect. ~2 we 
introduce the model used to study radial migration. In Sect. ~3 
we show the assumption used in the simulation. In Sect. ~4 we 
show the results that can be drawn from our calculations and 
finally the Sect. ~5 is devoted to the conclusions.

\section{Planets migration model}

\indent In a recent paper by Del Popolo et al. (1999), we studied how
dynamical friction, due to small planetesimals, influences the
evolution of KBOs having masses larger than
$10^{22}$ g. We found that the mean eccentricity of large mass particles
is reduced by dynamical friction due to small mass particles in 
timescales
shorter than the age of the solar system for objects of mass equal or
larger than $10^{23}$ g. Moreover the dynamical drag, produced by 
dynamical friction of objects of masses $\geq 10^{24}$ g, is 
responsible for the loss of angular momentum and the fall through 
more central regions in a timescale $ \approx 10^9$ yr. Here we 
apply a similar model to study the
planets radial migration. We suppose that a single planet moves in a
planetesimal disc under the influence of the gravitational force
of the Sun. The equation of motion of the planet can be written as:
\begin{equation}
{\bf \ddot r}= {\bf F}_{\odot}
%+ {\bf F}_{planets}+{\bf F}_{tide}+
%{\bf F}_{GCM}+{\bf F}_{stars}
+{\bf R}
%+{\bf F}_{other}
\end{equation}
(Melita \& Woolfson 1996), where the term ${\bf F}_{\odot}$ represents
the force per unit mass from the Sun, while ${\bf R}$ is the dissipative
force (the dynamical friction term-see Melita \& Woolfson 1996). 
%%In order to take into account dynamical friction, we need a 
%%suitable formula for a disc-like structure such as the 
%%protoplanetary disc. 
Calculations involving dynamical friction that are used to study 
planetesimal dynamics often use Chandrasekhar's theory 
(Stewart \& Wheterill 1988; Ida 1990; Lissauer \& Stewart 1992) for homogeneous and isotropic distribution of lighter particles.
% Tolto
%%If the velocity distribution is Maxwellian, the
%%dynamical
%%friction force can be written as:
%%\begin{eqnarray}
%%{\bf F} &= &-4\pi n m_1 (m_1+m_2) G^2
%%\frac{{\bf v_1}}{v_1^3}  \cdot \nonumber \\
%%&   &
%%\log \Lambda [erf(X)-2 X \exp(-X^2)/\sqrt(\pi)]
%%\label{eq:chac}
%%\end{eqnarray}
%%(Chandrasekhar \& von Neumann 1942; Chandrasekhar 1943; Binney \& 
%%Tremaine
%%1987) where n is the number density of field particles, $m_1$ is the 
%%mass
%%of the test particle, $m_2$ is the mass of a field one, $v_1$ the 
%%velocity
%%of the test particle, $\log{\Lambda}$ is the Coulomb logarithm and
%%$X=v_1/\sqrt{(2 \sigma)}$, being $\sigma$ the velocity dispersion.
%%Eq. (\ref{eq:chac}) cannot be used for non-spherically-symmetric 
%%systems.
%
This choice is not the right one, since dynamical friction in discs
differs from that in spherical isotropic
three dimensional systems. This is because disc evolution is influenced by effects different than those producing the evolution 
of stellar systems: \\
1) In a disc, the contribution to the friction coming from close 
encounters is comparable to that due to distant encounters
(Donner \& Sundelius 1993, Palmer et al. 1993). \\
2) Collective effects in a disc are much stronger than those in a 
three-dimensional system (Thorne 1968). \\
3) The peculiar velocities of planetesimals in a disc are small.
This means that differential rotation of the disc dominates over
planetesimals' relative velocities. \\
4) The velocity dispersion of particles in a disc potential is anisotropic. \\
%%
%%N-body simulations and observations show that
%%the radial component of the dispersion, $\sigma_R$, and the 
%%vertical one, $\sigma_z$, are characterized by a ratio 
%%$\sigma_{\rm R}/\sigma_{\rm z} \simeq 0.5$ for planetesimals 
%%in a Keplerian disc (Ida et al. 1993). The velocity 
%%dispersion evolves through gravitational scattering 
%%between particles. Gravitational scattering between particles 
%%transfers the energy of the systematic rotation
%%to the random motion (Stewart \& Wetherill 1988).
%%In other words, the velocity distribution of a Keplerian 
%%particle disc is ellipsoidal with ratio 2:1 between the 
%%radial and orthogonal (z) directions
%%(Stewart \& Wetherill 1988). The anisotropic velocity 
%%dispersion in discs implies that Chandrasekhar's formula is 
%%not appropriate for studying dynamical friction in them.\\
%%
\indent We assume that the matter-distribution is disc-shaped and that 
it 
has a velocity distribution described by:
\begin{equation}
n({\bf v},{\bf x})=n({\bf x})\left( \frac 1{2\pi }\right)
^{3/2}\exp \left[ -\left( \frac{v_{\parallel }^2}{2\sigma _{\parallel 
}^2}+%
\frac{v_{\perp }^2}{2\sigma _{\perp }^2}\right) \right] \frac 1{\sigma
_{\parallel }^2\sigma _{\perp }}
\end{equation}
(Hornung \& al. 1985, Stewart \& Wetherill 1988)
where $ v_{\parallel }$ and $\sigma_{\parallel}$ are the velocity and 
the velocity dispersion in the
direction parallel to the plane while  $ v_{\perp }$ and
$\sigma_{\perp}$ are those in the
perpendicular direction. We suppose that  $\sigma_{\parallel}$
and $\sigma_{\perp}$ are constants and that their ratio is simply taken
to be 2:1. Then according to Chandrasekhar (1968) and Binney (1977) 
we may write the force components as:
\begin{eqnarray}
F_{\parallel } &= & k_{\parallel}v_{1 \parallel}= B_{\parallel 
}v_{1\parallel }
\left[ 2\sqrt{2\pi } \overline{n} G^2\lg \Lambda  m_1m_2\left(
m_1+m_2\right)
\frac {\sqrt{1-e^2}}{\sigma _{\parallel }^2\sigma _{\perp
}}\right]\label{eq:b1}
\end{eqnarray}
\begin{eqnarray}
F_{\perp } &= &k_{\perp}v_{1 \perp}= B_{\perp
}v_{1_{\perp }}  \left[ 2\sqrt{2\pi } \overline{n} G^2\lg \Lambda m_1m_2\left( 
m_1+m_2\right)
\frac {\sqrt{1-e^2}}{\sigma _{\parallel }^2\sigma _{\perp 
}}\right]\label{eq:b2}
\end{eqnarray}
where
\begin{eqnarray}
B_{\parallel } &= &\int_0^\infty 
\exp{ \left[ -\frac{v_{1\parallel
}^2}{%
2\sigma _{\parallel }^2}\frac 1{1+q}-\frac{v_{1\perp }^2}{2\sigma
_{\parallel }^2}\frac 1{1-e^2+q}\right] } \times 
%% \nonumber\\ &   &
\frac {dq}{\left[ \left( 1+q\right)
^2\left( 1-e^2+q\right) ^{1/2}\right]}
\label{eq:b3}
\end{eqnarray}
\begin{eqnarray}
B_{\perp } &= &\int_0^\infty  
\exp \left[ -\frac{v_{1\parallel
}^2}{2\sigma
_{\parallel }^2}\frac 1{1+q}-\frac{v_{1\perp }^2}{2\sigma _{\parallel 
}^2}%
\frac 1{1-e^2+q}\right]  \times 
%% \nonumber \\&   &
\frac {dq}{\left[ \left( 1+q\right) \left(
1-e^2+q\right) ^{3/2}\right] }
\label{eq:b4}
\end{eqnarray}
and
%\begin{equation}
%e=\sqrt{(1-\frac{\sigma_{\perp}^2}{\sigma_{\parallel}^2})}
%\end{equation}
\begin{equation}
e=(1-\sigma_{\perp}^2/\sigma_{\parallel}^2)^{0.5}
\end{equation}
while $\overline{n}$ is the average spatial density,
$m_1$ is the mass
of the test particle, $m_2$ is the mass of a field one, and 
$\log{\Lambda}$ is the Coulomb logarithm. The frictional drag on 
the test particles may be written as:
\begin{equation}
{\bf F}=-k_{\parallel}v_{1 \parallel} {\bf e_{\parallel}}-
k_{\perp}v_{1 \perp}{\bf e_{\perp}}
\label{eq:dyn}
\end{equation}
where ${\bf e_{\parallel}}$ and ${\bf e_{\perp}}$ are two versors
parallel and perpendicular to the disc plane.\\
\indent
 When $B_{\perp}>B_{\parallel}$, the drag caused by dynamical 
friction
will tend to increase the anisotropy of the velocity distribution of
the test particles.
In other words, the dynamical drag experienced by an object of
mass $m_1$ moving through a less massive non-spherical
distribution of objects of mass $m_2$ is not directed in the direction
of the relative motion (as in the case of spherically symmetric 
distribution
of matter). Hence the already flat distribution of more massive objects 
will
be further flattened during the evolution of the system (Binney 1977).
As shown by Ida (1990), Ida \& Makino (1992) and Del Popolo et al. 
(1999)
damping of eccentricity and inclination is more rapid than radial 
migration
so in this paper we deal only with radial migration and we assume that
the planet has negligible inclination and eccentricity,
$i_{\rm p} \sim e_{\rm p} \sim 0$ and that the initial heliocentric distance of the planet is $5.2 {\rm AU}$.
The objects lying in the plane have no way
of knowing that they are moving into a non-spherically symmetric
potential. Hence we expect that the dynamical drag is directed
in the direction opposite to the motion of the particle:
\begin{equation}
{\bf F} \simeq -k_{\parallel}v_{\parallel} {\bf e_{\parallel}}
\end{equation}

\section{Simulation parameters}

In order to calculate the effect of dynamical friction on the orbital 
evolution of the planet, we suppose that
$\sigma_{\parallel}$=$2 \sigma_{\perp}$ and that the dispersion 
velocities
are constant. If the planetesimals attain dynamical equilibrium, their
equilibrium velocity dispersion, $\sigma_{\rm m}$, would be comparable 
to
the surface escape velocity of the dominant bodies (Safronov 1969) such 
that
\begin{equation}
\sigma_{\rm m} \sim v_{\rm esc} \sim
\left(\frac{G m_{\ast} }{ \theta r_{\ast}}\right)^{1/2}
%
%%\sigma_{\rm m} \sim v_{\rm esc} \sim \left(\frac{2G M_{\rm p} }{R_{\rm
%%p}}\right)^{1/2}
%
%\sim 0.1 \left(\frac{M_{\rm f}}{10^{23}{\rm g}}\right)^{1/3}
%\left(\frac{\rho_{\rm f}}{1{\rm g cm}^{-3}}\right)^{1/2}
%{\rm km/s}.
\label{eq:vesc}
\end{equation}
where $\theta$ is the Safronov number, 
%%$2<\theta<5$,
$m_{\ast}$ and $r_{\ast}$ are the mass and radius of the largest
planetesimals, 
(note that the planetesimals velocity dispersion, $\sigma_{\rm m}$,
now introduced, is the velocity dispersion to be used for
calculating the $\sigma$ which is present in the dynamical friction 
force).
If instead we consider a two-component system, consisting of one 
protoplanet and many equal-mass planetesimals 
%(we remember that planetesimals with a
%general mass distribution are well described by equal mass planetesimals
%with an effective mass $m_{\rm eff}$ (Ida \& Makino 1993),
the velocity dispersion of planetesimals in the neighborhood of the
protoplanet depends on the mass of the protoplanet. When the mass of the
planet, $M$, is $\le 10^{25}$ g, the value of $<e^2_{\rm m}>^{1/2}$
(being $e_{\rm m}$ the eccentricity of the planetesimals) is
independent of $M$ therefore:
\begin{equation}
e_{\rm m} \simeq 20 (2 m /3 M_{\odot})^{1/3}
\end{equation}
(Ida \& Makino 1993) where $m$ is the mass of the planetesimals. When 
the
mass of the planet reaches values larger than $10^{25}$-$10^{26}$ g at
1 AU, $<e^2_{\rm m}>^{1/2}$ is proportional to $M^{1/3}$:
\begin{equation}
e_{\rm m} \simeq 6 (M/3M_{\odot})^{1/3}
\end{equation}
(Ida \& Makino 1993).
As a consequence also the dispersion velocity in the disc is
characterized by two regimes being it connected to the eccentricity
by the equation:
\begin{equation}
\sigma_{\rm m} \simeq (e_{\rm m}^2+i_{\rm m}^2)^{1/2} v_{\rm c}
\end{equation}
where $i_{\rm m}$ is the inclination of planetesimals and
$v_{\rm c}$ is the Keplerian circular velocity.
The width of the heated region is roughly given by
$4 [(4/3)(e_{\rm m}^2+i_{\rm m}^2)a^2+12 h_{\rm M}^2 a^2]^{1/2}$ (Ida \& 
Makino 1993)
where $a$ is the semi-major
axis and $h_{\rm M}= (\frac{M+m}{3 M_{\odot}})^{1/3}$
is the Hill radius of the protoplanet. The increase in velocity dispersion
of planetesimals around the protoplanet decreases the
dynamical friction force (see Eq. \ref{eq:dyn}) and consequently 
increases the migration time-scale.\\
\indent In the simulation we assume that the planetesimals have all 
equal
masses, $m$, and that $m<< M$, $M$ being the planet mass. This 
assumption
does not affect the results, since dynamical friction does not depend
on the individual masses of these particles but on their overall 
density.
%As previously remarked in the previous section,
%since damping of eccentricity and inclination is more rapid
%than radial migration we assume that the planet has
%$i_{\rm p} \sim e_{\rm p} \sim 0$
%and that the initial heliocentric distance of the planet is $5.2 {\rm AU}$.
We also assume that the surface density in planetesimals varies
as $\Sigma(r)=~\Sigma_{\odot}(1 {\rm AU}/r)^{3/2}$,
where $\Sigma_{\odot}$, the surface density at 1 AU, is a free 
parameter.
The total mass in the planetesimal disc within radius $r$ is then:
\begin{equation}
M_{\rm D} \simeq 1.4 \times 10^{-3} (\Sigma_{\odot}/10^3 {\rm 
g/cm^2})(r/{\rm AU})^{0.5}
\end{equation}
We assume that $\simeq 1 \%$ of the disc mass is in the form of solid
particles (Stepinski \& Valageas 1996). To be more precise, assuming 
metal
abundance $Z=0.02$, the disc mass in gas interior to Jupiter's orbit
is $0.16 M_{\odot} (\Sigma_{\odot}/10^3 {\rm g/cm^2})$ (Murray et al. 
1998). In our model the gas is almost totally dissipated when the
planet begins to migrate and we assume that the value of disc mass 
reported
in the following part of the paper is contained within $40$ AU
(Weidenshilling 1977). We integrated the equations of motion in 
heliocentric coordinates using the Bulirsch-Stoer method.

\section{Results}

\subsection{Migration in a non-evolving disc}

\indent Our model starts with a fully formed gaseous giant planet of
$1 M_{\rm J}$ at $5.2$ AU.
As mentioned in the previous section, the circumstellar
disc is assumed to have a power-law radial density and to be 
axisymmetric.
According to several evidences showing that the disc lifetimes range 
from
$ 10^5$ yr to $10^7$ ~yr (Strom et al. 1993; Ruden \& Pollack 1991), 
we assume that the disc has a nominal effective lifetime of $10^6$ 
years (Zuckerman et al. 1995). This assumption refers to the gas disc. 
Usually, this decline of gas mass near stars is more rapid than the 
decline in the mass of orbiting particulate matter (Zuckerman et
al. 1995). Moreover the disc is populated by residual planetesimals for
a longer period. We are interested in studying the migration due to
interaction with planetesimals and for this reason we suppose that the
gas is almost dissipated when the planet starts its migration.
Since Jupiter-mass planets may require most of the lifetime of the disc
to accrete ($10^6$ to $10^7$ years) (and meanwhile the disc is subject 
to evolutionary changes) and since the disk is also subject to evolution
after this time interval (Pollack et al. 1996; Zuckerman et al. 1995), 
we incorporated this possibility in our model by running also models 
allowing some disc to dissipate during the planets migration.
We integrated the model introduced in the previous section for
several values of the disc surface density or equivalently several disc
masses: $M_{\rm D}=0.01$, 0.005, 0.001, 0.0005, 0.0001 $M_{\odot}$.\\
%%We remember that estimates of the "minimum mass" disc necessary to form
%%our planetary system range from $\simeq 0.001$ to $0.1 M_{\odot}$
%%(Weidenschilling 1977; Boss 1996).
%%Because of the loss of planetesimals ejected through orbital encounters 
%%with giant planets (see Murray et al. 1998), estimates
%%on the low end of this range may be insufficient to form our solar 
%%system; consequently, the minimum mass for our solar
%%system may be probably $\simeq 0.06 M_{\odot}$ (Boss 1996).
\begin{figure}
%\resizebox{\hsize}{!}{\includegraphics{f1.ps}}
\psfig{file=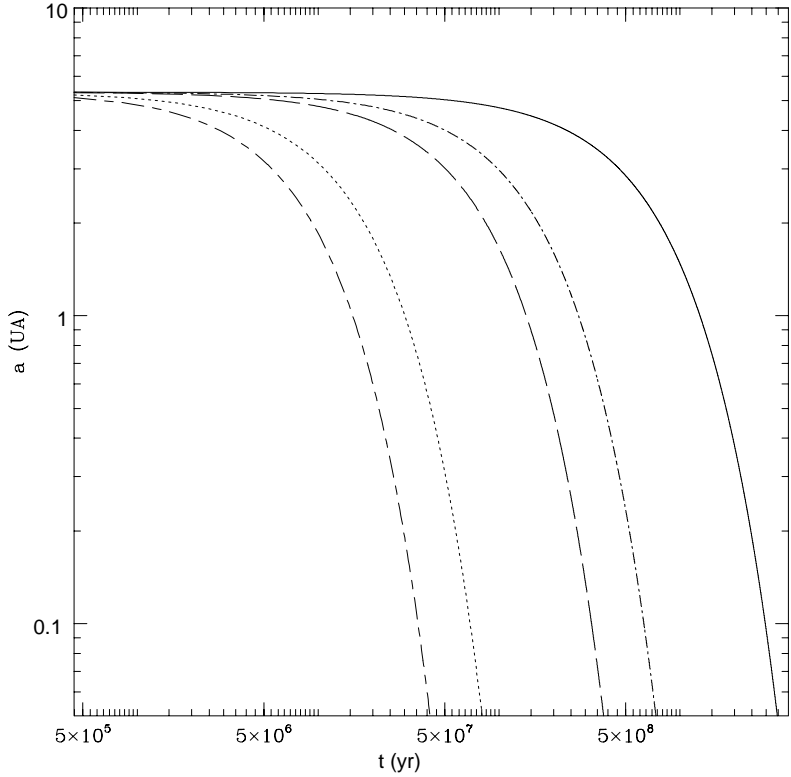,width=12cm}
%\picplace {2.0cm}
\caption[]{The evolution of the a(t) of a Jupiter-mass planet,
$M=1 M_{\rm J}$ in a disc with $\Sigma=\Sigma_{\odot} (1 {\rm 
UA}/r)^{3/2}$ for several values of the planetesimal disc mass,
$M_{\rm D}=0.01 M_{\odot}$ (short-dashed long-dashed line),
0.005 $M_{\odot}$ (dotted line), 0.001 $M_{\odot}$ (long dashed line),
0.0005 $M_{\odot}$ (dot-short dashed line) and 0.0001 $M_{\odot}$ 
(solid line). $\Sigma$ is supposed to remain constant in time.
}
\label{Fig. 1}
%and for $R_{\rm f}=3h^{-1}Mpc$.}
\end{figure}
%%Infrared observations of solar-mass T Tauri stars suggest that the total
%%mass in particulate matter at $r<100$ AU results in the range
%%$10^{-5}$-$10^{-2} M_{\odot}$ (Beckwith et al. 1990) while the total 
%%mass of solid  material in the present-day solar system is $\simeq 10^{-4} 
%%M_{\odot}$ (Stepinski \& Valageas 1996). \\
\indent The results of this first set of calculations (assuming that the
disc does not evolve) is shown in Fig. 1. The curves show the evolution 
of a 1 $M_{J}$ planet in a disc with planetesimals surface density
$\Sigma=\Sigma_{\odot} (1 {\rm AU}/r)^{3/2}$. The simulation is 
started with the planet at 5.2 AU and $i_{\rm p} \sim e_{\rm p} \sim 0$. 
%%%The assumption of zero
%inclination and eccentricity comes from several previous results
%%(Ida 1990; Ida \& Makino 1992; Del Popolo et al. ~1999)
%%%showing that in the case of objects $< M_J$ inclinations are damped in 
%%%a timescale much shorter than that of radial migration. 
In any case, similarly to what showed by
Murray et al. (1998), planets with masses $M>3 M_{\rm J}$ during their
migration can increase the value of $e_{\rm p}$. The curves correspond, from 
bottom
to top (short-dashed long-dashed line, dotted line, long dashed line,
dot-short dashed line, solid line) to the following values of $M_{\rm 
D}$:
0.01, 0.005, 0.001, 0.0005, 0.0001 $M_{\odot}$. As expected the most 
massive
disc ($0.01 M_{\odot}$) produces a rapid radial migration of the planet.
Discs having masses lower than $0.01 M_{\odot}$ produce a smaller radial
migration of the planet.
%and if the disc have mass $<10^{-4} M_{\odot}$ the migration does not 
%initiate.
In particular, we found that for $M_{\rm D}=0.01 M_{\odot}$ the
planet moves to $0.05$ AU in $\simeq 4 \times 10^7$ yr. The migration 
halts for the reason explained in the following
(with the term 'halt' we mean that the planet has not had time
to migrate any further, even if it is still migrating). If $M_{\rm D}=0.005, 0.001,
0.0005, 0.0001 M_{\odot}$ we have respectively, for the time needed to 
reach 0.05 AU:
$\simeq 8 \times 10^7$ yr, $\simeq 4 \times 10^8$ yr, $\simeq 7.5 \times =
10^8$ yr,
$\simeq 3.5 \times 10^9$ yr.
%$a=0.02, 1.7, 3, 4.7$.
\begin{figure}
%\resizebox{\hsize}{1!}{\includegraphics{f6.ps}}
\psfig{file=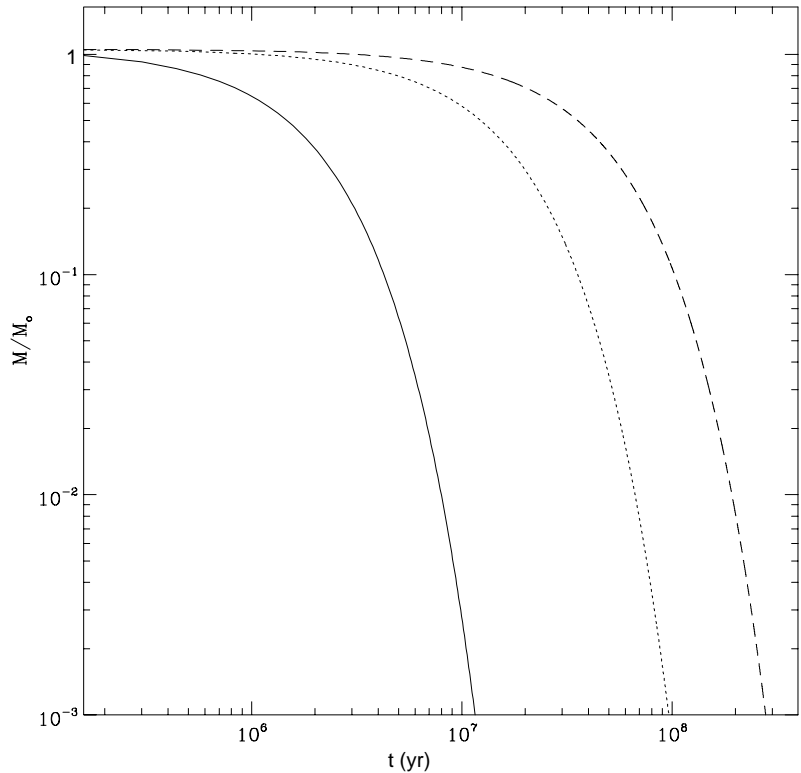,width=12cm}
%\picplace {2.0cm}
\caption[]{The assumed evolution of the planetesimals mass. The mass, 
from
the initial value $M_o$, reduces to $10^{-3} M_{\rm o}$, in $\simeq 
10^7$ yr
(solid line), $\simeq 10^8$ yr (dotted line), $\simeq 3 \times 10^8$
(dashed line)}
\label{Fig. 2}
%and for $R_{\rm f}=3h^{-1}Mpc$.}
\end{figure}
In other words, disc mass is one of the parameters that controls radial
migration. An interesting feature of the model is that migration 
naturally
halts without needing any peculiar mechanisms that avoid the planet from
plunging into the central star. In fact as shown in Fig. 1 the migration 
time
to reach $\simeq 0.05$ AU increases with decreasing disk mass. If the 
disk
mass is $\le 0.00008 M_{\odot}$ the time needed to reach the quoted 
position
is larger than the age of the stellar system and the planet does not 
fall
into the star. Then, the planet can halt its migration without falling 
in
the star if the initial disc mass is $\le 0.00008 M_{\odot}$.
%%or
%%if, due to disc evolution, the disc
%%mass falls below the quoted value (see next section).
Even if the disc density does not fall below the critical value, the 
planet
must halt at several $R_{\ast}$ from the star surface
($R_{\ast}$ is the stellar radius). In fact solid
bodies
cannot condense at distances $\leq 7 R_{\ast}$,
%%
%%and existing planetesimals, whose
%%orbits might evolve to smaller radii,
and planetesimals cannot
survive for a long time at distances $\leq 2 R_{\ast}$. When the planet
arrives at this distance the dynamical friction force switches off
and its migration stops. This means that the minimum
value of the semi-major axis that a planet can reach is $\simeq 0.03$ 
AU.\\

\subsection{Migration in an evolving disc}

\indent In the previous calculations, we supposed that the disc mass did 
not undergo time evolution but in reality the disc evolves as the planet 
moves inward and tends to dissipate. \\
%%\indent A major unsolved problem for protoplanetary discs is determining 
%%the mechanism through which such discs evolve.
\begin{figure}
%\resizebox{\hsize}{!}{\includegraphics{f4.ps}}
\psfig{file=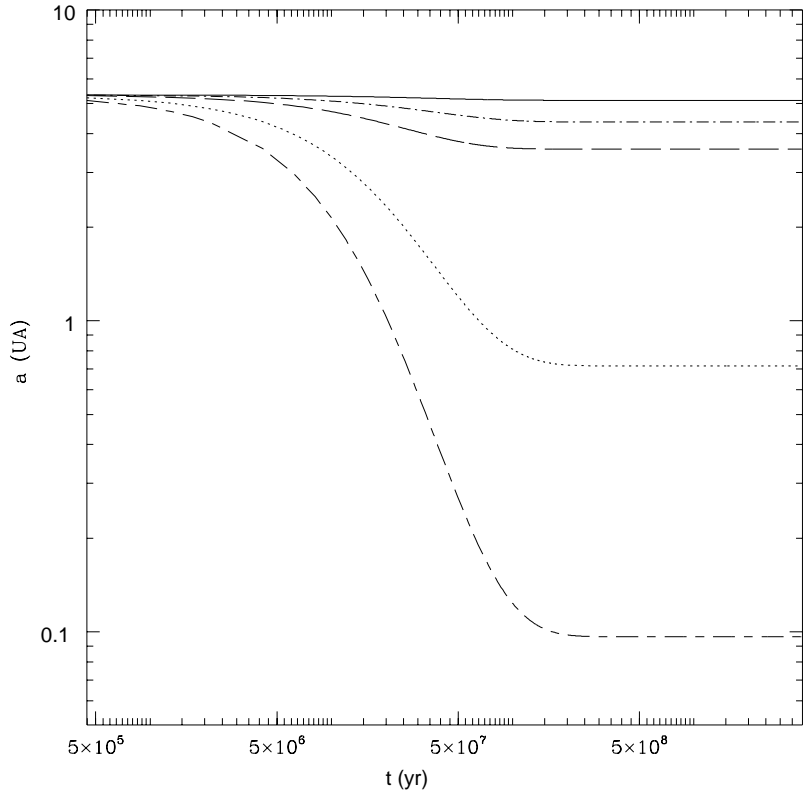,width=12cm}
%\picplace {2.0cm}
\caption[]{Same as Fig.1 but now we suppose that the disc mass decreases
exponentially in $3 \times 10^8$ yr (see Fig. 2). }
\label{Fig. 3}
%and for $R_{\rm f}=3h^{-1}Mpc$.}
\end{figure}
%%%To begin with gas and solid particles follow a different evolution.
%%The expectation is that
%%%Disc evolution should produce inward mass transport and outward angular
%%%momentum transport (Pringle 1981), yet recently it has been showed that, 
%%%in some situations, convective instability leads to turbulent stresses that
%%%operate in the reverse sense and transport angular momentum inward (Ryu 
%%5\& Goodman 1992; Kley et al. 1993). Gravitational torques associated 
%%%with nonaxisymmetric density distributions can result in rapid
%%%disc evolution (Larson 1984; Boss 1984). A third possible source of disc
%%%evolution are magnetic fields. Strom et al. 1993 presented evidences
%%%for inner disc lifetimes ranging from $\simeq 10^5$ to $\simeq 10^7$ yr,
%%%which are consistent with estimates for the evolution times of viscous 
%%%accretion disc models (e.g. Ruden \& Pollack 1991) and observational evidence 
%%%that some circumstellar discs dissipate after $10^6-10^7$ yr (Zuckerman et al. 
%%%1995).
\begin{figure}
%\resizebox{\hsize}{!}{\includegraphics{f2.ps}}
\psfig{file=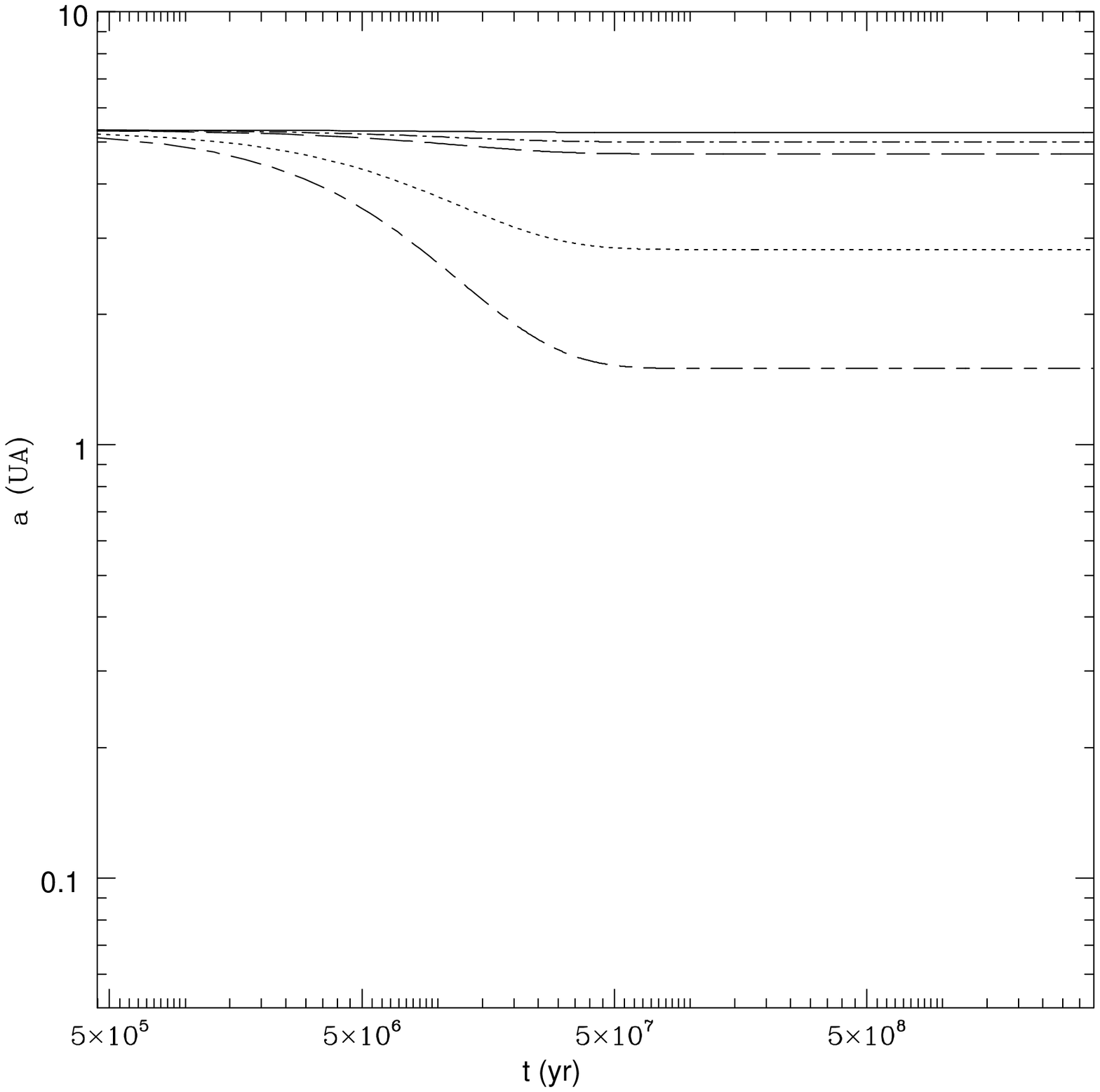,width=12cm}
%\picplace {2.0cm}
\caption[]{Same as Fig. 2 but with planetesimals mass evolving in
$10^8$ yr (see Fig. 2)}
\label{Fig. 4}
%and for $R_{\rm f}=3h^{-1}Mpc$.}
\end{figure}
In our model, we are fundamentally interested in the evolution of solid
matter and planetesimals in discs. To this aim, it is very important to 
note that the distribution of solid particles follows a global time 
evolution, which accompanies
%%, but is non identical
%%to, the global
the time evolution of the gaseous component of the disc. Due to viscous 
torques, the gaseous disc spreads and its mass diminishes.
%%due to viscous torques.
If initially the solid particles are small and coupled to the gas, they
decouple from it when they gain mass because of coagulation.
%%
%%Initially, the solid particles are very small, so
%%they are coupled to the gas. With time the solid particles gain mass 
%%due
%%to the process of coagulation and decouple from the gas, following an
%%evolution different from that of the gas
(Stepinski \& Valageas 1996).
Particles having radius $r < 0.1$ cm, can be considered perfectly 
coupled to the gas, while those having $r>10^5$ ~cm can be considered 
completely decoupled
from it and their mean velocities remain practically unchanged with time
(Stepinski \& Valageas 1996). Moreover, as shown in a recent study by
Ida et al. (2000) the radial migration of a planet of
Jupiter-mass produce a very rapid capture of planetesimals in the 2:1
and 3:2 resonances: the resonance capture occurs if the migration
time, $\tau_{\rm mig}$, of the planet is $\tau_{\rm mig} > 10^4$ yr
for 2:1 resonance and if $\tau_{\rm mig} > 10^3$ yr for 3:2 resonance. 
If the result is correct this means that the disc should be rapidly 
depleted with a consequent rapid stopping of migration. As can be 
understood by what previously told, disc evolution depends on disc 
and system characteristics. \\
%% For example the nearby classical T Tauri star DH Tau exhibits excess 
%% emission at 10 $\mu$m which is consistent with predictions based on 
%% circumstellar disc models. The T Tauri star DI Tau identified by 
%% Skrutskie et al. (1990) on the basis of 12 $\mu$m IRAS data, 
%% as an object in the process of dissipating its circumstellar disc, is
%% found to have no infrared excess at a wavelength of 10 $\mu$m.
\begin{figure}
%\resizebox{\hsize}{!}{\includegraphics{f3.ps}}
\psfig{file=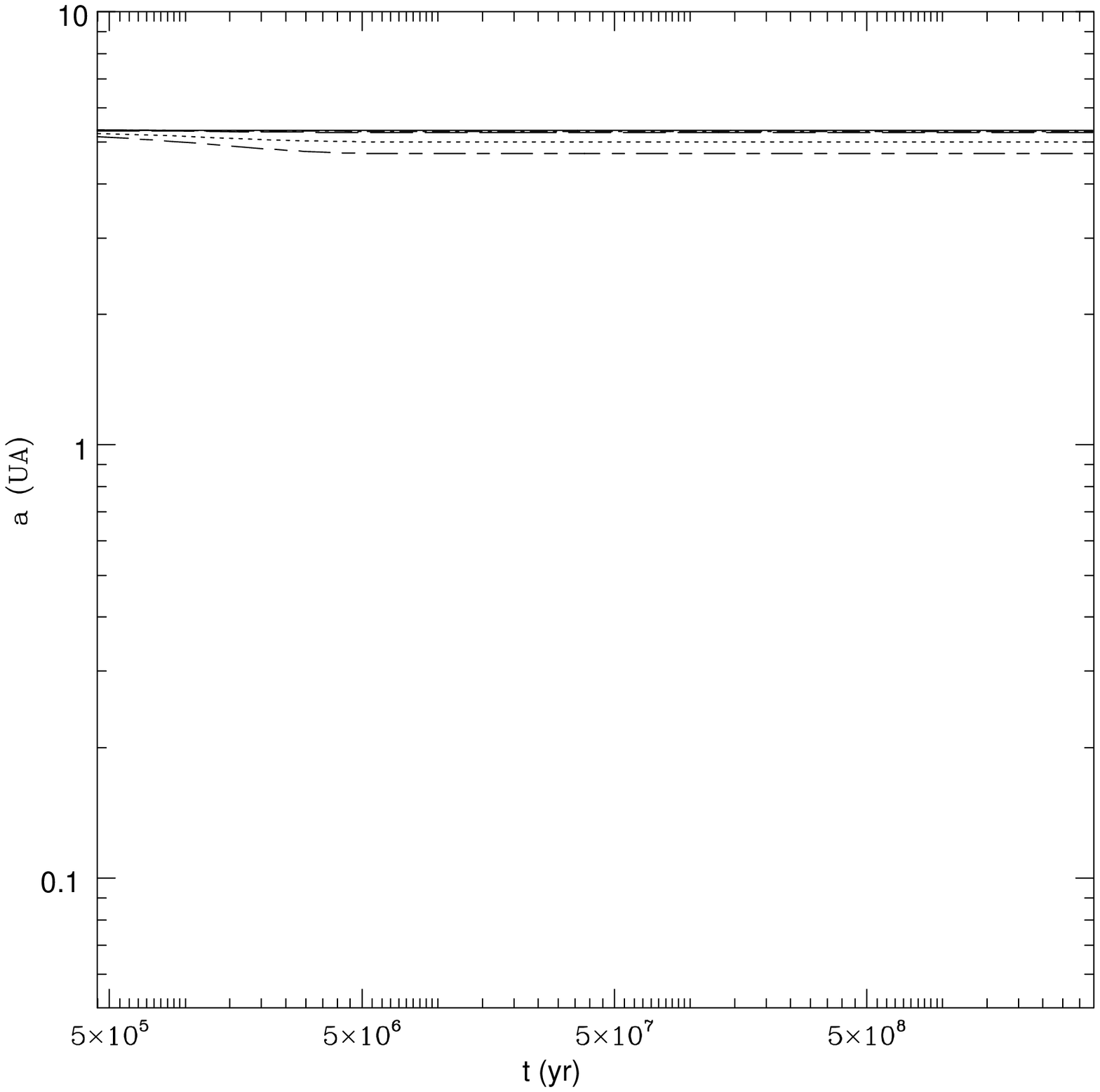,width=12cm}
%\picplace {2.0cm}
\caption[]{Same as Fig. 4 but with planetesimals mass evolving in $10^7$ 
yr
(see Fig. 2)}
\label{Fig. 5}
%and for $R_{\rm f}=3h^{-1}Mpc$.}
\end{figure}
%% While both objects appear to have the same stellar mass, age, and 
%% rotation rate, they differ in two fundamental respects:
%% DH Tau is a single star with an active accretion disc and DI Tau is a 
%% binary system lacking such a disc. So the formation of a sub-stellar mass
%% companion has probably led to the rapid evolution of the circumstellar 
%% disc of DI Tau.\\
\indent In this paper we tried to take account of disc evolution by 
supposing
that the total mass in solids decays with time from its original value
to the present value as shown in Fig. 2
%%for solar system $10^{-4} M_{\odot}$ or lower values
(see Stepinski \& Valageas 1996). The results
of this calculation are shown in Fig. 2 to Fig. 5. \\
\indent Fig. 2 shows the evolution of the disc mass used in the 
calculations of radial migration (see Stepinski \& Valageas 1996, 
Fig 6). We suppose that the mass in the disc decreases 
exponentially with time from its original value $M_o$
to $10^{-3} M_o$ in $3 \times 10^8$ yr (dashed line),
$10^8$ yr (dotted line) and $10^7$ yr (solid line). Fig. 3 is 
the same as Fig. 1 but now we suppose that the mass
in the disc decreases as described.
%As shown, the situation is not very different from that seen in Fig. 1.
%If $M_{\rm D}=0.01$ the planet reaches $a=0.05$ in $\simeq 5 \times 
%10^7$ yr

%while if $M_{\rm D}=0.005, 0.001, 0.0005, 0.0001$
%we have $a=0.13, 2.48, 3.59, 4.83$.
If $M_{\rm D}=0.01 M_{\odot}$, the planet stops its migration at $a 
\simeq 0.1$ AU,
while if $M_{\rm D}=0.005, 0.001, 0.0005, 0.0001 M_{\odot}$,
we have $a \simeq 0.7, 3.5, 4.3, 5$ AU.
Fig. 4 is obtained by supposing that the mass decreases, as previously
quoted, in $10^8$ yr.
As can be shown the planet embedded in a disc having
$M_{\rm D}=0.01$, 0.005, 0.001, 0.0005, 0.0001 $M_{\odot}$ 
respectively
migrates to 1.47, 2.7, 4.58, 4.88, 5.1 AU.
Finally in Fig. 5 the time-scale for disc evolution is $10^7$ yr.
The planet embedded in a disc with $M_{\rm D}=0.01$, 0.005, 0.001, 
0.0005, 0.0001
$M_{\odot}$ migrates respectively to 4.58, 4.88, 5.13, 5.17, 5.2 AU. \\
\indent Our distribution of final masses and heliocentric distances 
predicts
that massive planets can be present at any heliocentric distances 
between
their formation locations and extremely small orbits, depending on the
initial mass of the disc and its evolution. As we shall show in the
following, the model can explain the locations of not only the close
companions (at $\leq 0.1$ AU), but it can also reproduce other observed
planets, including Jupiter. 
%Configuration of planets like $\tau$ Bootis b, 51 Peg b, having very 
%small
%semimajor axis,  can be reproduced by high values of the initial disc 
%mass
%and low time evolution (e.g., $M_{\rm D}=0.01 M_{\odot}$, $t_{\rm 
%evol}=10^8$ yr)
%(see also Boss 1996) and a disc mass a bit lower can
%explain the configuration of planets like 55 Cnc b
%($a=0.11$ AU), $\rho$ CrB b ($a=0.23$ AU). The parameters of 47 UMa =
%b
%($a=2.11$ AU) can be explained supposing a low mass disc (e.g.,
%$M_D=10^{-3} M_{\odot}) or disc evolution (e.g., $M_D=10^{-2} =
%M_{\odot},
%$t_{evol}=10^8$ yr)
\begin{figure}
%\resizebox{\hsize}{!}{\includegraphics{fig5.ps}}
\psfig{file=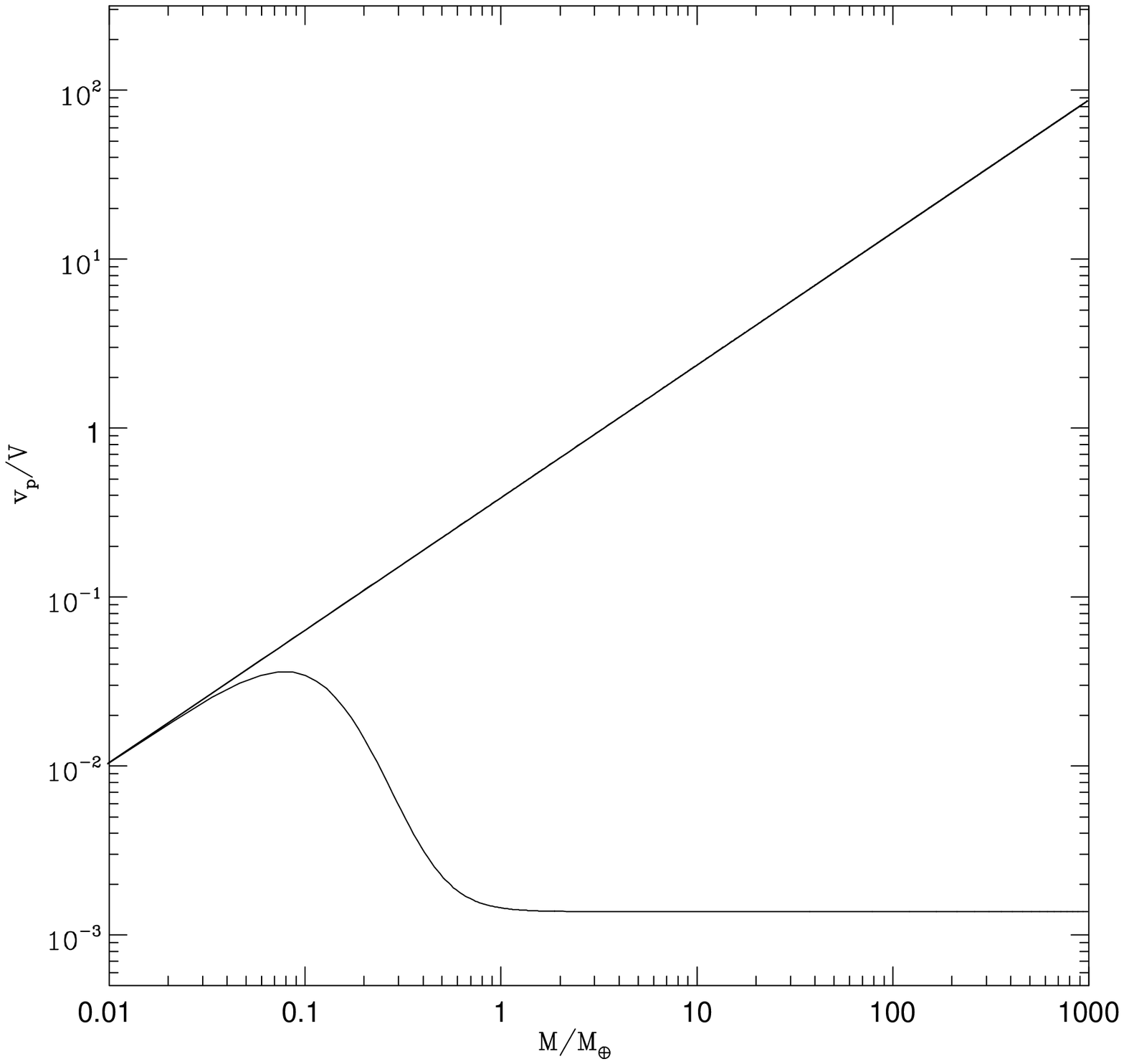,width=12cm}
%\picplace {2.0cm}
\caption[]{Drift velocity, $\frac{d r}{d t}$, as a function of mass.
Velocities are normalized to $V=2 \frac{M_{\oplus}}{M_{\odot}}
\frac{\pi \Sigma r^2}{M_{\odot}} (\frac{r \Omega}{\sigma})^3 r \Omega$
where $M_{\oplus}$ is an Earth mass, $\Sigma$ the surface density, 
$\Omega$
is the angular velocity and $\sigma$ the dispersion velocity. The 
assumed
conditions are those considered appropriate for the Jovian region and
assuming that $M_{\rm D}=0.01 M_{\odot}$. The line $\propto M$ 
correspond to model
%% $type$ I drift 
described by Ward (1997)
%%. $Type$ I 
and its behavior is valid till
$\simeq 0.1 M_{\oplus}$ but beyond this value there is a transition to
a behavior $\propto M^0$}. 
%%($type$ II motion)}.
\label{Fig. 6}
%and for $R_{\rm f}=3h^{-1}Mpc$.}
\end{figure}

\subsection{Comparison of the model results with observations}

\indent Configuration of planets like $\tau$ Bootis b, 51 Peg b, 
having very small semimajor axis,  can be reproduced by models with 
no evolution (see Fig. 1) or high values of the initial disc mass
and low time evolution (e.g., $M_{\rm D}=0.01 M_{\odot}$,
$t_{\rm evol}=4 \times 10^8$ yr)(Boss 1996) and disc mass a bit 
lower can explain the configuration of planets like 55 Cnc b
($a=0.11$ AU), $\rho$ CrB b ($a=0.23$ AU). The parameters of 47 UMa 
b ($a=2.11$ AU) can be explained, for example, by supposing a low 
mass disc and evolution (e.g.,
$M_{\rm D}=5 \times 10^{-3} M_{\odot}$, $t_{\rm evol}=3 \times 
10^8$ yr).
%or disc evolution (e.g., $M_{\rm D}=10^{-2} M_{\odot},
%$t_{evol}=10^8$ yr)
Planets like 70 Vir b ($a=0.43$, $e_{\rm p}=0.4$; 
$M \sin{i_{\rm p}} \sim 6.6 M_J$) and
HD 114762 b ($a=0.3$; $e_{\rm p}=0.25$;
$M \sin{i_{\rm p}} \sim 10 M_{\rm J}$) have
high eccentricities and masses larger
than Jupiter's. The low semimajor axis can be explained by radial 
migration,
as shown, while the high value of eccentricities can be explained in a 
model
of interaction planet-planetesimals like ours in a way similar to that 
shown
by Murray et al. (1998). If a planet having mass $M \ge 3 M_{\rm J}$, 
which is
the case of 70 Vir b and HD 114762,
moves in a planetesimal disc during interactions,
planetesimals scattered from their Hill sphere can be ejected with
$|\frac{\Delta E}{\Delta L}|<1$, (where $\Delta E$ and 
$\Delta L$ are respectively the energy and angular momentum
removed from a planet by the ejection of a planetesimal),
and the eccentricity $e_{\rm p}$ tends to
increase. For sake of completeness we must say that there are 
also some systems a bit puzzling. For example $v$ And has 
$M \sin{i_{\rm p}}=0.68$ and an high eccentricity, 
$e_{\rm p} \sim 0.15$, HD210277 has $M \sin{i_{\rm p}}=1.36$ and
$e_{\rm p}=0.45$. Such high eccentricities could be explained 
supposing an encounter with an object having mass $\simeq M_{\oplus}$ 
or alternatively supposing that the systems are seen at small $i_{\rm p}$.
%%
%%could arise from an encounter with an object
%%of a few $M_{\oplus}$ or their masses may be larger
%%because the systems are seen at small $i$.
In the case of the companion to 16 Cyg B ($M \sin{i_{\rm p}}=1.5$; 
$e_{\rm p} \sim 0.68$) the high eccentricity may be due to 
interactions with the stellar companion (Murray et al. 1998).\\
%%to
%%16 Cyg B.\\
\indent Our Solar System could have been subject
%%not have been immune
to giant planet migration. For example, a shrinkage of the orbit 
of Jupiter of 0.1-0.2 AU could naturally explain the depletion of 
the outer asteroid belt (Fernandez \& Ip 1984; Liou \& Malhotra 1997). 
Some of our model runs produce a 1 $M_J$ planet that move from 5.2 AU 
inwards for a fraction of AU. The planetesimal disc enabling this 
small migration has a lifetime $\simeq 10^7$ yr, so that the disc 
gas must have disappeared soon after Jupiter fully formed 
%%soon after Jupiter grew to its
%%present mass,
%%well before Jupiter could spiral into the Sun
(Boss 1996). \\
\indent This last result is obviously strictly valid only for a 
single planet orbiting around the Sun because in presence of several 
planets, migration becomes more complex. A close example is that of 
the solar system. In this case two planets, Uranus and Neptune were 
subject to outward migration, which is the opposite of what expected. 
Several models have been proposed to explain this outward migration. 
A first model is connected to gravitational scattering between
planet and residuals planetesimals (Malhotra 1993; Ida et al. 2000).
%% If a planetesimal in a near-circular orbit similar to that of
%% the planet is ejected into a Solar system escape orbit, the planet suffers
%% a loss of orbital angular momentum and a corresponding change of
%% orbital radius. Conversely, planetesimals scattered inwards would cause
%% an increase of orbital radius and angular momentum of the planet.
%% A single massive planet scattering a population of planetesimals
%% in near-circular orbits in the vicinity of its own orbit would suffer
%% no net change of orbital radius as it scatters approximately
%% equal numbers of planetesimals inwards and outwards. However in
%% some peculiar situations, such as that encountered in the region
%% of Jovian planets,
%% things go differently from this picture
%% (Fernandez \& Ip 1984). In particular, as Jupiter preferentially
%% removes the inwards scattered Neptune planetesimals, the planetesimal
%% population encountering Neptune at later times is increasingly
%% biased towards objects with specific angular momentum larger than Neptune's.
%% Encounters with this planetesimal population produce a net gain of angular
%% momentum, hence an increase in its orbital radius.\\
A second model allowing Neptune outward migration is connected to 
%% $type$ II drift.
the dissipation in the protostellar nebula.
In this case both inwards and outwards planet migration are allowed.
In fact in a viscous disc, gas inside a particular radius,
known as the radius of maximum viscous stress,
$r_{\rm mvs}$, drifts inwards as it loses angular momentum while gas outside
$r_{\rm mvs}$ expands outwards as it receives angular momentum
(Lynden-Bell \& Pringle 1974). Neptune's outwards migration is due to
the fact that the gas in the Neptune forming region has a tendency to
migrate outwards (Ruden \& Lin 1986).
%%
%%For example, in the solar system, Uranus and
%%Neptune migrate outward rather than inwards because Jupiter acts like 
%%an
%%inner absorbing boundary similar to a nearby stellar surface.
%%Since the
%%outer planets push the inner ones toward the star, migration begins
%%at lower values of the disc density while migration of massive planets
%%produces a depletion of the disc. This means that it is unlikely that
%%two massive planets migrate to short-period orbits. \\
%% \indent 
In Fig. 6 we show the drift velocity, $\frac{d r}{d t}$, as function
of mass, $M$. As shown, objects having masses $ < M_{\oplus}$ have 
velocity drift increasing as $M$, while after a threshold mass any 
further mass increases begins to slow down the drift. As the threshold 
is exceeded
the motion fairly abruptly converts to a slower mode in which the
drift velocity is independent of mass. As previously explained, this
behaviour is due to the transition from a stage in which the dispersion
velocity is independent of $M$ to a stage in which it increases with 
$M^{1/3}$ (Ida \& Makino 1993). This last stage is known as the 
protoplanet-dominated stage.
%%\begin{figure}
%%%\resizebox{\hsize}{1}{\includegraphics{am.ps}}
%%\psfig{file=f6.ps,width=12cm}
%\picplace {2.0cm}
%%\caption[]{The evolution of the planetesimals mass. The mass
%%reduces to $10^{-3}$ its initial value, $\rho_o$ in $\simeq 10^7$ yr
%%(solid line), $\simeq 10^8$ yr (dotted line), $\simeq 3 \times 10^8$
%%(dashed line)}
%%\label{Fig. 6}
%and for $R_{f}=3h^{-1}Mpc$.}
%%\end{figure}
The phenomenon is equivalent to that predicted in the density wave 
approach
(Goldreich \& Tremaine 1980; Ward 1997). In this approach, the density 
wave torques
repel material on either side of the protoplanet's orbit and attempt to
open a gap in the disc. Only very large objects are able to open and
sustain the gap. After gap formation, the drift rate of the planet is
set by disc viscosity and is generally smaller than in absence of the 
gap.
We stress that the decaying portion of the curve
corresponding to
the transition from the first to the second stage does not correspond to
any particular model because following Ida \& Makino (1993) we do not 
have information on the evolution of $\sigma$ in the transition regime.
%% We also stress that the behaviour $\propto M$ is valid in the $type$ I
%% regime for $M <0.1 M_{\odot}$. 
\subsection{Enhancements of metallicity}

\indent Another important point is that the dynamical processes leading to
planets migration can also affect the evolution of the central star.
Gonzales (1997, 1998a,b) showed that several stars with short-period
planets have high metallicities, $[{\rm Fe/H}] \geq 0.2$. Gonzales 
(1998b) proposes that their metallicities have been enhanced by the 
accretion of high Z-material which leads to the speculation that 
there may be a relationship between stars with higher metallicities 
and stars with planets. Alternatively, the correlation could arise, 
if metal-rich stars have metal-rich discs which are more
likely to form planets.\\
\indent Several mechanisms have been proposed to explain the high 
metallicity of stars having extra-solar planets. One of this mechanism 
is related to the Lin et al. (1996) migration model but this
%% in which several multiple Jupiter-mass planets migrate inwards 
%% as a result of a $type$ II migration. Today, we are
%% only able to observe the last of the planets which did not spiral in
%% their parent star. Since a Jupiter-mass planet is made of roughly 10
%% $M_{\oplus}$ cores of chondritic composition (Anders \& Grevesse 1989), 
%% it would take $\simeq 5$ such Jupiter-mass planets to raise the 
%% stellar metallicities to those observed in 51 Peg b. The 
model has two severe drawbacks.
%% 1) improbability of such kind of systems; \\
%% 2) after the migration of the first Jupiter-mass planets, the disc 
%% should be
%% depleted and no other planet could fall in the star. Then the material
%% 'delivered' by a single Jupiter should not be enough to change the star
%% composition.\\
Models in which gas disc material accretes on the star are not 
able to significantly alter the observed metallicity because the 
disc has a metallicity slightly larger than the star. 
%% \indent 
Good results are obtained in models in which asteroids or 
planetesimals accrete on to the star (Murray et al. 1998). As 
previously quoted, Murray et al. (1998) suggest that a giant planet can
induce eccentricity growth among residual planetesimals through resonant
interactions. Subsequent close encounters cause most of the affected
planetesimals to be ejected outwards while the planet migrates
inward. A substantial population of planetesimals could induce a
Jupiter-mass planet to migrate a large distance inward. Neglecting
any planetesimals that are scattered into the star until the planet 
reaches
its final orbit and assuming that the fraction of planetesimals 
scattered onto
the star consist of only those that become planet crossing before 
colliding with the
star, the mass accreted on the star is given by:
%In the model, the mass accreted on the star
%(taking account only of planetesimals planetesimals which become planet
%crossing before colliding with the star)
%is given by:
\begin{equation}
M_{\rm acc} \sim f(a) M/\alpha \label{eq:mu}
\end{equation}
where
\begin{equation}
f(a) \simeq 0.15 (a/{\rm AU})^{-0.374} \label{eq:mu1}
\end{equation}
(Ford et al. 1999), where $f(a)$ is the fraction
of planetesimals scattered onto the star by the planet at distance $a$ 
and
$\alpha$ is a parameter $\simeq 0.5-1$. As shown by
Eq. \ref{eq:mu1} the fraction of planetesimals scattered onto the star 
increases
with decreasing distance from it. At $a=0.05$ $f(a) \sim 0.46-0.92$ if
$\alpha=0.5-1$.
For example, the mass of planetesimals which would be scattered into
51 Peg, whose final orbit has $a=0.05$ AU,
%or $\rho^1$ Cnc, whose final orbit has $$ are respectively,
is $130 M_{\oplus}$ (Ford et al. 1999).
%and $180 M_{\oplus}$.
By starting from a star with solar metallicity adding  $130 M_{\oplus}$ 
of
asteroids to 51 Peg the observed $[{\rm Fe/H}]$ increases to 0.48
%while adding
%$180 M_{\oplus}$ to $\rhp^1$ Cnc increases $ [{\rm Fe/H}]$ to 0.39
(Ford et al. 1999). The value found is an inferior limit because
it takes account only of planetesimals scattered from the planet located 
in
its final orbit. Moreover it is calculated for a value of
$\Sigma_{\odot} \sim 4000 {\rm g/cm^2}$ less than the maximum disc mass
used in Murray et al. (1998). Using the largest values of 
$\Sigma_{\odot}$
used by Murray et al. (1998) one expects a value of $M_{\rm acc}$ double 
than
that previously quoted.\\
\indent Even if we assume $M_{\rm acc} \sim 130 M_{\odot}$, this value 
is
larger than that observed in 51 Peg ($ [{\rm Fe/H}] \sim 0.21$).
Then following the Ford et al. (1999) model for 51 Peg, the prediction
for metallicity abundance is larger than that observed. This means that
unless most of the acquired
heavy elements are able to diffuse in the radiative interior, the
planetesimal-scattering scenario for orbital migration
would require more than 90 \% of the close encounters to
result in the outward ejection of planetesimals (Sandquist et al. 
1998).\\
\indent Our model predicts a smaller value for metallicity close
to the observed value for 51 Peg.
%%The problem of metallicity overabundance of Murray et al. (1998) model
%%is not present in ours.
In fact, our model, similarly to that by Murray et al. (1998), explains 
the
planets migration by planet-planetesimals interaction, but differently 
from
the Murray's et al. (1998) our model needs less planetesimal mass for 
radial
migration.
%%In
%%any case, also in our model when a Jupiter-mass planet migrates, 
%%planetesimals
%%are also scattered.
When the planet reaches its final location, Eq. \ref{eq:mu} $ \div $\ref{eq:mu1}
(giving the quantity of planetesimals scattered in the star) can be 
applied
%, taking account that our disc is less massive and than
after scaling Murray et al. (1998) result to reflect our disc mass.
In the case of
the more massive disc ($M_{\rm D}=0.01 M_{\odot}$)
we find $M_{\rm acc} \sim 40 M_{\oplus}$ and
$[{\rm Fe/H}] \sim 0.2$. \\
\indent An important point to stress is that the plausibility of such
an explanation depends, among other, on the size of the stellar
convective envelope at the time of accretion. In order to be efficient,
the accretion must take
place sufficiently late in the stellar evolution when the outer 
convective
envelope is shallow. Accretion taking place while the star was still on 
the pre-main-sequence, and consequently having a large convective 
envelope, would have little effect on stars' observed metallicities. 
The time it takes for a planet to migrate to a $<0.1$ AU orbit, in 
our model, is $>10^7$ yr.
Once the planet stops its migration, planetesimals inside its orbit
are quickly cleared out. As shown by Ford et al. (1999), the right 
time to produce the observed metallicities is at $2-3 \times 10^7$ yr. 
This value can therefore be regarded as an important constraint for 
disc models.
%%%%Moreover, as we previously quoted,
%%%%a Jupiter-mass planet needs $10^6-10^7$ yr in order for it to 
%%%%complete
%%%%formation and in
%%%%this phase velocity dispersion in the disk are small.
%%%%For example, a big planetesimal of
%%%%$10^{24}$ g falls in the star, from some AU in $\simeq 10^7$ yr.
%In other terms, during
%the period neceessary for planet formation falls in the star $\simeq 
%10^6$
%planetesimals of the quoted mass................
%%%%This planetesimals mass also contribute to change the metallicity 
%%%%abundance
%%%%of the star.
\section{Conclusions}

\indent The discovery (Mayor \& Queloz 1995; Marcy \& Butler 1996;
Butler \& Marcy 1996; Butler et al. 1997; Cochran et al. 1997; 
Noyes et al. 1997) of extra-solar planets has revitalized the 
discussion on the theory of planetary system formation and evolution. 
Although close giant planets formation may be theoretically possible
(Wuchterl 1993; Wuchterl 1996), it requires the initial formation of a 
solid core of at least 5 $ \div$ 10 $M_{\oplus}$ which may be difficult 
to achieve very close to the parent star. It is therefore more likely 
that Jupiter-mass extra-solar planets cannot form at small heliocentric 
distances (Boss 1995; Guillot et al. 1996).
After those discoveries, the idea that planets can migrate radially
(Goldreich \& Tremaine 1980; Ward \& Hourigan 1989; Lin \& Papaloizou 
1993; Lin et al. 1996) for long distances has been taken more seriously 
than was made in the past years. \\
\indent In this paper, we showed that
dynamical friction between the planet and a planetesimals disc
is an important mechanism for planet migration.
%is due to dynamical friction between the planet and a planetesimals 
%disc.
We showed that migration of $1 M_J$ planet to small heliocentric 
distances ($0.05 $ AU) is possible for a disc with a total mass of 
$10^{-4} \div 10^{-2} M_{\odot}$
(we remember that, according to Stepinski \& Valageas 1996,
and Murray et. al 1998,  only  $\simeq 1 \%$ of the disc mass is in
the form of solid particles) if the planetesimal disc does not dissipate
during the planet migration or if the disc has $M_{\rm D} >0.01 
M_{\odot}$ and the planetesimals are dissipated in $\sim 10^8$ yr.
The model predicts that massive planets can be present at any 
heliocentric distances for the right value of disc mass and 
time evolution. \\
\indent We also showed that the drift velocity of planets and than the 
migration time are very similar to the predictions of the density 
wave approach (Ward 1997): the drift velocity increases as $M$ for 
masses smaller than $0.1 M_{\oplus}$ and is constant for larger 
masses. Finally we showed that the metallicity enhancement observed 
in several stars having extrasolar planets can also be explained, 
similarly to what proposed by Murray et al. (1998) and Ford et al. 
(1999) by means of scattering of planetesimals onto the parent star, 
after the planet reached its final configuration.
Comparing our model with other models, that attempt to explain 
planets migration, we think our model has some advantages: \\
1) differently from models based on the density wave theory (Goldreich 
\& Tremaine 1980; Ward 1986, 1997), our model does not require a peculiar
mechanism to stop the inward migration (Lin et al. 1996). Planet halt
is naturally provided by the model. It can explain planets found at
heliocentric distances of $> 0.1$ AU or planets having larger values of
eccentricity. It can explain metallicity enhancements observed in stars
having planets in short-period orbits;\\
2) differently from Murray et al. (1998) model, our model shows that
radial migration is possible with not too massive planetesimals
disc (which is one of the drawbacks of the Murray's et al. (1998) model) 
and predicts the right metallicity enhancement. 
%\begin{flushleft}
%{\it Acknowledgements}
%\end{flushleft}
\section*{Acknowledgments}
%%%%%%Work partially supported by funds ex-60\% 98.
%%V.A.-D. would like to thank Prof. Giuseppe Moncada.
We thank the anonymous referees whose comments and suggestions 
helped us to improve the quality of this work. We are grateful to 
E. Ford and E. Spedicato for stimulating discussions during the 
period in which this work was performed.
A. Del Popolo and E.Nihal Ercan would like to thank
Bo$\breve{g}azi$\c{c}i University
Research Foundation for the financial support through the project code
01B304.

%\end{acknowledgements}
                           
%\begin{thebibliography}{999}

\end{document}